# Compressive sensing based velocity estimation in video data

Ana Miletić, Nemanja Ivanović
Faculty of Electrical Engineering
University of Montenegro
ana_miletic@t-com.me, nemanjaivanovic@gmx.com

*Abstract* - **This paper considers the use of compressive sensing based algorithms for velocity estimation of moving vehicles. The procedure is based on sparse reconstruction algorithms combined with time-frequency analysis applied to video data. This algorithm provides an accurate estimation of object's velocity even in the case of a very reduced number of available video frames. The influence of crucial parameters is analysed for different types of moving vehicles.**

## I. INTRODUCTION

What the Shannon-Nyquist sampling theorem teaches us is that sampling must be performed at least two times as fast as the signal's actual bandwidth in order to preserve the information it carries. In applications such as surveillance, an increased sampling density either leaves us with too many samples that must be compressed in order to be transmitted or is very expensive to implement. Sampling at lower rates would be much easier to carry out. Besides, it would also require less storage space. As the number of surveillance cameras grows, the amount of data that has to be transported grows. This leads to a higher risk of a data congestion which may prevent proper object detection. Thus, it would be much more appropriate for each camera to transmit a small amount of data which carries just enough information for the tracking. Several methods in the literature are developed to tackle the earlier mentioned sampling issues using compressive sensing (CS) [1]-[8]. In video applications, these approaches [6]-[8] provide precise velocity estimation from a diminished set of randomly selected video frames. The missing frames are a discontinuity in video streaming. They could be a result of possible reductions of storage requirements or some unwanted circumstances. The data loss produces large errors during the velocity estimation, as it appears as a certain kind of noise [5]. Generally, the whole CS concept is based on the possibility of reconstructing a signal from a small subset of available samples [1]. In order to achieve this, the signal needs to fulfill the condition of sparsity, which means that only a certain amount of transform coefficients could be considered as non-zeros [1], [2].

The theoretical background on the IF estimation based on the time-frequency analysis is presented in Section II. The procedure for motion parameters estimation is described in Section III, while the employed algorithm is briefly explained in Section IV. The experimental results and comparisons are given in Section V.

## II. INSTANTANEOUS FREQUENCY AND ITS ESTIMATION USING TIME-FREQUENCY ANALYSIS

The estimation of motion parameters is usually performed through spectral analysis methods and techniques [12]-[14]. In the case of constant velocity, the Fourier transform can normally be used, but this is not the case in practical applications, where nonstationary signals are used with varying spectral content. In other words, these signals can be observed as sinusoidal waves whose frequencies vary with time. The most significant parameter in the analysis of these signals is their instantaneous frequency (IF) [9]. It is a parameter which defines the position of the signal's spectral peak. In video processing, it resembles the object's current velocity in the domain of time. In order for the velocity to be estimated, the IF needs to be calculated first, usually based on the time-frequency analysis methods. The commonly used are [9]:

- The Short-time Fourier transform (STFT), where $p(t)$ is multiplied by a window function $w(t)$ and the Fourier transform is calculated as the window slides along the time axis:

$$STFT(t,\omega) = \sum_{\tau} w(\tau) p(t+\tau) e^{-j\omega\tau} \qquad (1)$$

- The spectrogram, the square module of the STFT which is used in many practical applications:

$$SPEC(t,\omega) = \left| STFT(t,\omega) \right|^2 \qquad (2)$$

- The S-method (SM), is a time-frequency distribution that can be calculated using the STFT and improves results over the spectrogram [2]:

$$SM(t,\omega) = \sum_{i=-L}^{L} STFT(t,\omega+i\Theta) STFT^*(t,\omega-i\Theta) \qquad (3)$$

with STFT* being the complex conjugate of the STFT and 2L+1 being the width of the window function.

## III. ESTIMATION OF MOTION PARAMETERS IN VIDEO SEQUENCES

In order to estimate the velocity of an object in a 3D signal, the mentioned methods need to be combined with other advanced techniques, such as the SLIDE (subspace-based line detection) algorithm [15]. According to it, every video frame is firstly projected onto the coordinate axes, where the obtained projections are then used to produce an FM signal. The motion parameters can be calculated once the parameters of FM signal are known. A certain video frame that appears at time instant *t* can be represented as [12],[13],[15]:

$$F(x,y,t) = e(x,y) + o(x-x_0-v_x t, y-y_0-v_y t) \quad (4)$$

where $o(x,y)$ is the representation of the moving object, $e$ is the background, $(x_0,y_0)$ are the coordinates of the object's initial position and $(v_x,v_y)$ is its velocity. The next step is to calculate the frames' projections onto the coordinate axes. Since the same rules apply for both x and y axis, we can concentrate on only one for the sake of simplicity. Therefore, after summing (5) along y axis and applying basic operations, we obtain:

$$FP(x,t) = E(x) + O(x-x_0-v_x t) \quad (5)$$

Assuming that the background is steady, we may make the calculating process easier by deriving (6) with respect to t:

$$\frac{\partial FP(x,t)}{\partial t} = FP(x-x_0-v_x t) \quad (6)$$

In order to map this 2D function into the frequency domain, the µ-propagation can be used:

$$s(t) = \sum_x FP(x-x_0-v_x t)e^{j\mu x} \quad (7)$$

where the IF of $s(t)$, calculated based on µ, correlates with the object's velocity. The parameter µ refines any sudden velocity drops or rises and therefore contributes to the velocity estimation significantly. Its huge importance will additionally be discussed and shown through practical examples.

Once all the frames are available, these calculations are not difficult to perform. But if we are left with just a subset of frames, it is not simple to reconstruct the signal and to provide precise results. In other words, a subset $B(x,y,T)$ of the total set of frames $F(x,y,t)$ is observed. The lengths of $B$ and $F$ are $M$ and $N$, respectively, where $N \gg M$. In other words, instead of the whole signal $s(t)$, we are dealing with only a short vector of $M$ measurements taken at time instances $T = \{T_1, T_2, ..., T_M\}$, labeled as $s(T)$.

In order for velocity to be estimated, the STFT needs to be calculated first. Therefore, for each time instant $T_i$ (each windowed signal part), the measurement vector is formed as:

$$m(T_k) = w(\tau)s(T_k+\tau) \quad (8)$$

As the number of samples is decreased, so is the quality of the standard (initial) STFT of $m(T_k)$ which is no longer suitable for analysis, so it is necessary to recover the missing samples on the basis of the available ones [8]. This is the reason we apply compressive sensing reconstruction to each vector $m(T_k)$. In the equations that follow, the notation $T_k$ will be omitted for the sake of simplicity. Still, the same procedure is repeated for each time instant. According to the CS algorithms, we can write [1]:

$$m = \Phi d \quad (9)$$

where $d$ is the full set windowed µ-propagation vector, viewed as an $Nx1$ column vector, while $\Phi$ is the random Gaussian measurement matrix producing the measurement vector $m$ given in (8). Then, knowing that any signal in $R^N$ can be represented in Fourier basis using weighting coefficients $F_k$, we may also say that:

$$d = \sum_{k=1}^{N} \Psi_k F_k = \Psi F \quad , \quad (10)$$

where F is a $Nx1$ column vector of weighting coefficients and $\Psi$ is a full rank $NxN$ matrix. It is clear that $d$ and $F$ are the representations of the same signal, with $d$ being the time-domain representation and $F$ being the spectral representation. By combining previous equations, we have:

$$m = \Phi d = \Phi \Psi F = \Theta F \quad (11)$$

The aim of the whole process is to reconstruct $d$ (or its spectral equivalent $F$) from the set of $M$ measurements (the incomplete signal $m$). To that end, we need to solve a system of $M$ linear equations with $N$ unknowns. As $N>M$, there are fewer equations than unknowns, and the solution rely on optimization algorithms, such as $l_1 - norm$ minimization:

$$\min \left\| \tilde{F} \right\|_{l_1} \quad s.t. \quad m = \Theta \tilde{F}. \quad (12)$$

These minimizations are solved using different optimization algorithms [4],[5],[10],[11]. For each time instant the reconstructed vector $\tilde{F}$ corresponds to the STFT column. Then the SM is calculated in order to improve the efficiency of IF estimation [8].

## IV. THE ALGORITHM APPLICATION IN VEHICLES TRACKING

This section provides a short, step-by-step explanation of the Matlab algorithm used in [8] to obtain the velocity estimations. The main steps are:

- Divide the video into frames, and convert the frames to grayscale.
- A projection vector is made by calculating the differences between available frames and summing all difference images by columns.
- The µ-propagation vector is calculated.
- The initial STFT is calculated using the Hanning window based on a set of available samples. One of the main parameters in this algorithm is the window width ($N_p$). The initial versions of SPEC and SM are computed as well.
- The reconstructed version of STFT and SM are calculated by performing CS reconstruction during the calculation of STFT for each windowed part.

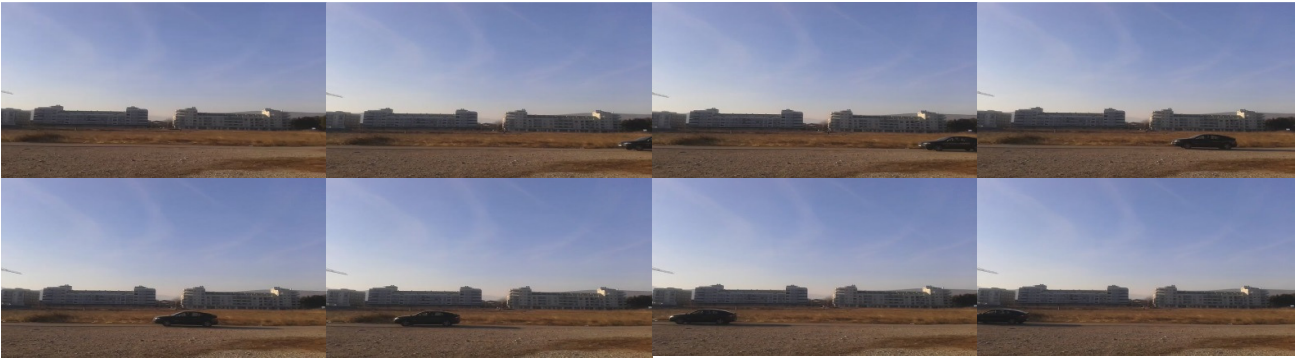

**Fig. 1. Video sequence frames: 1,39,49, 59,78,93,101 and 106**

The algorithm is applied to the velocity estimation of moving vehicles (cars, tracks, planes, etc.), in order to show that it can be efficiently used for different moving objects, without significant variations of parameters. Primarily, we observe the μ-parameter, which has a high impact on the overall results.

## V. EXPERIMENTAL RESULTS

*Example 1:* The first example considers a sequence which features an accelerating car. It is illustrated by a few frames in Fig. 1. The whole sequence consists of 121 frames (640 by 480 pixels). We estimate the car's velocity throughout the whole sequence based on 66 available frames (approx. 54%). The CS based SM and SPEC are shown in Fig. 2, where the SM outperforms the SPEC. The parameters μ and $N_p$ have been chosen in order to get as precise results as possible and equal 0.15 and 64, respectively. Ideally, the velocity variations at the beginning (frames 0-20) and at the end (frames 100-121) should not appear, but they are consequence of windy weather and camera trembling. In spite of that, the obtained results are quite good and clearly show the car is gradually accelerating. The influence of the parameter μ has been analyzed and the results are shown in Fig. 3.

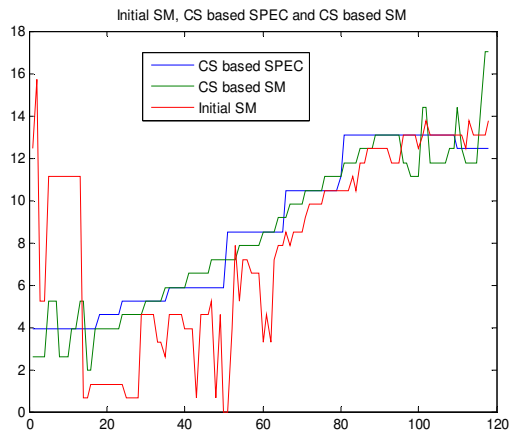

**Fig. 2. Velocity estimation: Initial SM (red line), CS based SPEC (blue line), CS based SM (green line) using $N_p$=64 and μ=0.15**

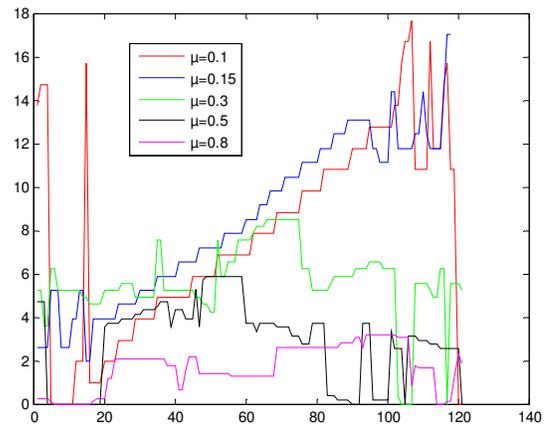

**Fig. 3. Velocity estimation using CS based S-method using $N_p$=64 and μ=0.1; μ=0.15; μ=0.3; μ=0.5; μ=0.9**

The velocity estimation is tested for different values of the parameter μ, showing that the optimal μ will always provide the smoothest changes in the vehicle velocity. The assumption is that the car cannot produce a fast velocity change within a couple of frames and the optimal μ can be selected as the line with the smoothest changes. Similar performance is obtained with other moving vehicles as well, with the conclusion that in most observed cases, optimal μ is between 0.1 and 0.3 (step 0.05 provides sufficient precision).

*Example 2:* Our second example illustrates a landing plane. It has been divided into 130 frames (each 640 by 360 pixels), 8 of which are shown in Fig.4. The velocity estimation is based on 70 available frames (53%). As shown in Fig. 5, the CS based SM again outperforms the SPEC which cannot track the velocity changes properly. In this example we used $N_p$=64 and μ=0.25. Note that the same window width can be used in all experiments, while the parameter μ is just slightly changed in this case compared to the previous example.

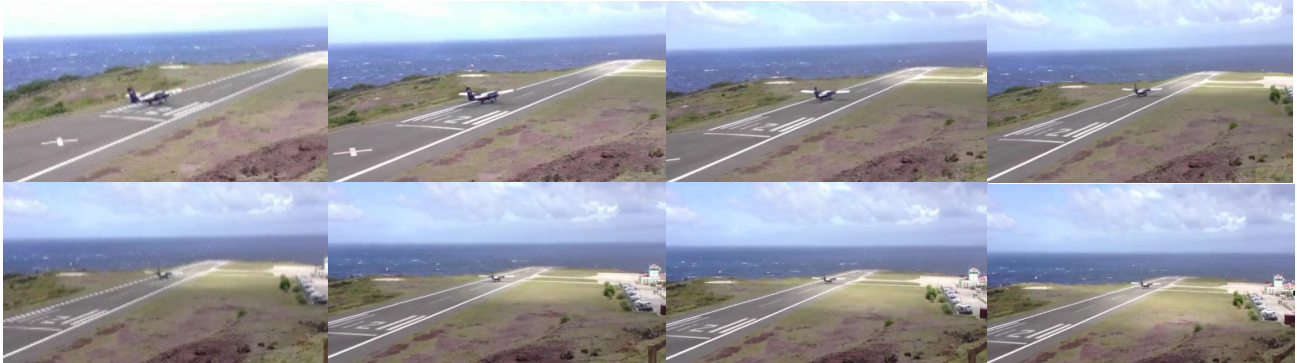

**Fig. 4. Video sequence frames: 1, 15, 32, 53, 74, 97,110 and 130**

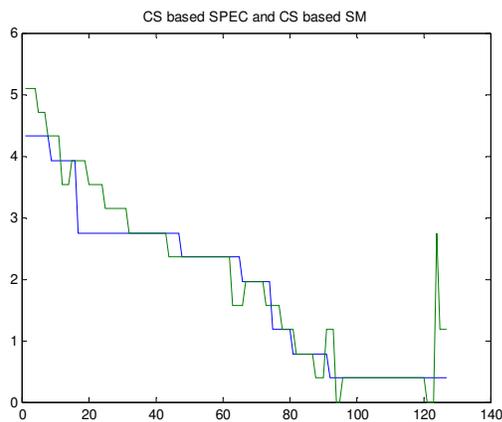

**Fig. 5. Velocity estimation using CS based spectrogram (blue line) and CS based S-method (green line) using $N_p=64$ and $\mu=0.25$**

## VI. CONCLUSION

The paper provides the analysis of CS based velocity estimation of vehicles using an incomplete set of frames. Particularly, the possibility of using CS based SM for estimating the motion parameters of different vehicles is examined. The influence and choice of µ-propagation parameter is examined as well, showing that for different moving objects the parameter changes just slightly, while the optimal value can be chosen as the smoothest velocity line.

### ACKNOWLEDGEMENT

The authors are thankful to Professors and assistants within the Laboratory for Multimedia Signals and Systems, at the University of Montenegro, for providing the ideas, codes, literature and results developed for the project CS-ICT (funded by the Montenegrin Ministry of Science).